\documentclass[a4paper, 12pt, openbib ]{article}
\usepackage{times}
\usepackage{pifont}
\usepackage{amsmath}
\usepackage{amssymb}
\usepackage{amsfonts}
\usepackage{graphicx}
\usepackage{floatflt}
\usepackage{geometry}
\usepackage{layout}
\usepackage{enumerate}
\usepackage{bm}
\usepackage{tikz}
\usetikzlibrary{patterns}
\def\ner{\boldsymbol}
\usepackage{amsthm}
\usepackage{thmtools}
\def\tfract#1/#2{{\textstyle{\raise0.8pt\hbox{$\scriptstyle#1$}\over%
\hbox{\lower0.8pt\hbox{$\scriptstyle#2$}}}}}
\def\mezzo{\tfract 1/2 }

\def\radi2k{\tfract 1/{\sqrt {2k}} }
\def\der{\partial }
\def\cvd{\vbox{\hrule \hbox to 9 pt {\vrule height 9 pt \hfil \vrule} \hrule}}
\def\downnormalfill{$\,\,\vrule depth4pt width0.4pt
\leaders\vrule depth 0pt height0.4pt\hfill\vrule depth4pt width0.4pt\,\,$}
\def\WT#1{\mathop{\vbox{\ialign{##\crcr\noalign{\kern3pt}
      \downnormalfill\crcr\noalign{\kern0.8pt\nointerlineskip}
      $\hfil\displaystyle{#1}\hfil$\crcr}}}\limits}

\def\be{\begin{equation}}
\def\ee{\end{equation}}
\def\bes{\begin{equation*}}
\def\ees{\end{equation*}}
\def\bea{\begin{eqnarray}}
\def\eea{\end{eqnarray}}
\def\beas{\begin{eqnarray*}}
\def\eeas{\end{eqnarray*}}
\def\ba{\begin{array}{rcl}}
\def\ea{\end{array}}
\def\der{\partial}
\numberwithin{equation}{section}
\def\go{\leavevmode \raise.3ex\hbox{$\scriptscriptstyle \langle\!\langle\!  $}%
~\ignorespaces}
\def\gf{\relax \ifhmode \unskip~\else \leavevmode \fi \raise.3ex\hbox{$\! \scriptscriptstyle\rangle\!\rangle\, $}}
\title{
{\Large  \bf Critical angular velocity for vortex lines formation}
{\vskip 0.6 truecm}} 

\author{{\large  Enore~Guadagnini} \\  {\normalsize {~}} \\  {\normalsize  Dipartimento di Fisica {\it E. Fermi} dell'Universit\`a di Pisa,} \\  {\normalsize and INFN Sezione di Pisa,} \\ {\normalsize   Largo B. Pontecorvo  3, 56127 Pisa, Italy.} 
}
\date{}

\begin{document}

\maketitle 

\vskip 0.7 truecm 

\begin{abstract}

For helium II inside a rotating cylinder, it is proposed that the formation of  vortex lines of the frictionless superfluid component of the liquid is caused by the presence of the rotating quasi-particles gas. By minimising the free energy of the system, the critical value $\Omega_0$ of the angular velocity for the formation of the first vortex line is determined.  This value  nontrivially depends on the temperature, and numerical estimations of its  temperature behaviour are produced. It is shown that the latent heat for a vortex formation and the associated discontinuous change in the angular momentum of the quasi-particles gas determine the slope of $\Omega_0 (T)$ via some kind of Clapeyron equation. 

 \end{abstract}

\vskip 1.2 truecm

\section{Introduction}

The formation of vortex lines, with quantized vorticity, in helium II can be understood as a 
macroscopic quantum mechanics effect \cite{AM,RD}. Vortex lines in helium II have  been observed in  rotating containers   \cite{PS,YGP,MAM,PBRD,BEW} and,  by means of an analysis on friction and drag on quantised vortices \cite{SON,FRN}, some of their phenomenological coefficients have been deduced.  The formation of vortices has been studied \cite{BDV} also in the case of freely rotating fluid drops of helium II. Discussions on the possible connections of the quantised vortex dynamics with superfluid turbulence can be found for instance in \cite{CBAR} and in the articles collection of Ref.\cite{MOSC}.  

The emergence of vortex lines   is a common feature  of quantum liquids. Vortices have an important impact on  high-temperature superconductors \cite{TAN} and have been observed \cite{ZZZ,HE3,YYY}  in rotating superfluid $^3$He.    Arrays of  vortex lines have also been  observed \cite{RC2,ROC,ARVK, FRA,FCS2} and studied \cite{ELPS,FFF,DSI,AAQD,AFAS,BAY,EKK} in Bose-Einstein condensates of cold atoms. 
 
In order to investigate the mechanism of vortex formation,  in the present article the coming into being  of  a single vortex line in helium superfluid, which is  contained in a rotating cylinder,  is considered.   The determination of the  critical value $\Omega_0$ of angular velocity at which the first vortex line appears is  discussed. Since the  superfluid component of liquid  helium  has no viscosity and its dynamics  is unaffected by   the moving walls of the container, it is proposed that the creation of a vortex line is induced by the rotating quasi-particles gas.  Minimization of the free energy of the system is used to derive the value of $\Omega_0$, which turns out to  depend nontrivially on the temperature. The angular momentum and the  thermodynamic variables which characterise the formation of the vortex line are examined, and their statistical mechanics expressions are  determined.  It is shown that the latent heat for a vortex formation and the associated discontinuous change in the angular momentum of the quasi-particles gas determine the slope of $\Omega_0 (T)$ via some kind of Clapeyron equation. 

In order to make this article self-contained, a few basic definitions of helium superfluid and the main properties  of the quasi-particles  are briefly recalled in  Section~2, where a  derivation of the densities of the thermodynamic potentials energy, free energy and entropy is presented. The deduction of the critical value of the angular velocity for the formation of a  vortex line is contained in Section~3. In Section~4 it is shown that the latent heat of vortex formation and the discontinuous change in the angular momentum of the quasi-particles gas determine the slope of the curve $\Omega (T)$ by means of an equation of the Clapeyron type.  During the vortex formation, the changes in entropy and in angular momentum are computed. Finally, numerical estimations of the temperature dependence of $\Omega (T)$ are reported. The main conclusions are collected in Section~5.

\section{Superfluid and quasi-particles}

When the value of the temperature is below the critical value $T_0 \simeq 2.18$~K  corresponding to the $\lambda$-point, at ordinary pressure the behaviour of helium $\hbox{He}^4$ is similar (but it is not equal) to the behaviour of  a two-components liquid in which 

\begin{itemize}
\item one component, which  has velocity $\ner v_s$ and mass density $\rho_s$, corresponds  to  the so-called superfluid motion; this fluid component has no viscosity and carries zero entropy; 

\item the second component,  with velocity $\ner v_n$ and mass density $\rho_n$, corresponds to the normal motion and  behaves as a normal viscous fluid. 
\end{itemize}
 
\noindent This peculiar quantum liquid can be described \cite{L1,F1,K,WC, RJDO} by means of a gas of quasi-particles,  which represent  the localized energy fluctuations of the system above its ground state, and by means of additional degrees of freedom which are related with the global (zero entropy) motion of the ground state wave function, that will simply be called  the global motion of the condensate with macroscopic velocity $\ner v_s$.  The  mass density of the liquid helium  is given by 
\be
\rho = \rho_s + \rho_n \; , 
\label{2.1}
\ee
 and the momentum density is written as  
\be
\ner P / V = \rho_s \ner v_s + \rho_n \ner v_n \; . 
\label{2.2}
\ee
Let us consider an inertial reference system.  
When the condensate is at rest ($\ner v_s=0$) and when $\ner v_n =0$,  the dependence of the energy $\varepsilon $ of a single quasi-particle on its momentum $\ner p $ is given by the energy spectrum  $\varepsilon ( p)$, where $p = | \ner p |$.  For small momenta, the function $\varepsilon (p)$ has a typical linear behaviour, $\varepsilon \simeq u p $,  where $u$ denotes the speed of first sound.  In a neighbourhood of  $p_0 \simeq 2 \times 10^8 \, {\hbar / {\hbox {cm}}}$, the function $\varepsilon (p)$ has a deep local minimum and it can be approximated as $\varepsilon (p) \simeq \Delta + (p-p_0)^2 / 2 m^* $.  Quasi-particles  obey the Bose-Einstein statistics and the quasi-particles gas has vanishing chemical potential. 

\subsection{Quasi-particle energy spectrum}
 
The hydrodynamic motions of the superfluid  component and of the normal component of the liquid appear to be  essentially independent, apart from a modification of the energy spectrum of the quasi-particles which takes place when the relative velocity $\ner v = \ner v_n - \ner v_s$ is not vanishing. Let us consider a small portion of the liquid with well-defined thermodynamics variables and given velocities $\ner v_s$ and $\ner v_n$.  When  $\ner v = \ner v_n - \ner v_s \not= 0$, the energy spectrum of a single quasi-particle (which belongs to this part of the liquid)  with momentum $\ner p $   is given by 
\be
E_v (\ner p )=  \varepsilon (p) - \ner v \ner p = \varepsilon (p) - (\ner v_n - \ner v_s) \ner p \; . 
\label{2.3}
\ee
This equation is a consequence  \cite{L1}  of the nonrelativistic transformation properties of energy and momentum under a change of an  inertial reference system into another inertial reference system.  The peculiar form (\ref{2.3}) of the energy spectrum is responsible \cite{L1} of the absence of viscosity of the superfluid motion. 

Equation (\ref{2.3}) also determines the values of the local densities of the thermodynamic potentials which are specified by the rules of  statistical mechanics. For a portion of liquid, in thermal equilibrium with temperature $T$ and with well defined velocities $\ner v_n$ and $\ner v_s$,    the relevant thermodynamic potentials  ---that will be useful for the following discussion--- will now be computed. Let us concentrate on the case in which the values $ v_n = | \ner v_n|$ and $v_s= | \ner v_s|$ are smaller than any intrinsic velocity scale of helium liquid (speeds of the first and second sound,...), so that the thermodynamic potentials can be approximated by their Taylor expansion up to second  order   in powers of $\ner v_n$ and $\ner v_s$.  The low density approximation for the quasi-particles gas will also be considered, and thus the interactions between the quasi-particles will be neglected.

\subsection {Momentum density}

When the condensate is at rest ($\ner v_s =0 $), a small  portion of the liquid, in which $\ner v_n \not= 0$, has a momentum density $\ner P / V $ which is associated to the normal component  of the fluid  exclusively,  $\ner P / V = \rho_n \ner v_n $. This momentum density coincides with the momentum density of the quasi-particles gas,  $\ner P / V = \left [ \ner P / V \right ]_{q.p.} $. Let $n(\varepsilon )$ denote the quasi-particles (Bose-Einstein) distribution  function, at a given temperature $T\,$;  then 
\be
\left [ \ner P / V \right ]_{q.p.} = \int d \tau \, \ner p \, n (\varepsilon - \ner v_n \ner p) \, \simeq \, \ner v_n  \int d \tau \, (p^2 / 3 ) \, \left [ - {\der  n (\varepsilon ) \over  \der \varepsilon } \right ] = \ner v_n \, \rho_n \; , 
\label{2.4}
\ee
where  $ d \tau = d^3p / h^3$, and  a  first order expansion of $n (\varepsilon - \ner v_n \ner p) $ in powers of the velocity $\ner v_n$ has been considered.  In the integration function, the multiplicative factor $\ner p_i \ner p_j$ has been replaced by $\delta_{ij} \, p^2/3$. Equation (\ref{2.4}) determines \cite{L1} the value $\rho_n$ of the mass density of the  normal component  of the fluid 
 \be
 \rho_n =  \int d \tau \, (p^2 / 3 ) \, \left [ - {\der  n (\varepsilon ) \over  \der \varepsilon } \right ] \; . 
 \label{2.5}
 \ee
Let us now consider another small part of the liquid in which both velocities $\ner v_n$ and $\ner v_s$ are nonvanishing. In this case, the total momentum density is the sum of the momentum density $\left [ \ner P / V \right ]_{q.p.}$ due to the quasi-particles and the momentum density $ \left [ \ner P / V \right ]_{s}$ which is associated with the condensate motion with velocity $\ner v_s$, 
 \be
 \ner P / V = \left [ \ner P / V \right ]_{q.p.} + \left [ \ner P / V \right ]_{s} \; . 
 \label{2.6}
  \ee
In agreement with expressions (\ref{2.3}), the quasi-particles contribution is given by 
\bea
\left [ \ner P / V \right ]_{q.p.} &=& \int d \tau \, \ner p \, n (\varepsilon - (\ner v_n - \ner v_s ) \ner p) \nonumber \\
&\simeq& (\ner v_n - \ner v_s ) \int d \tau \, (p^2 / 3 ) \, \left [ - \der  n (\varepsilon ) / \der \varepsilon \right ] \nonumber \\
&=& \ner v_n \, \rho_n - \ner v_s \, \rho_n = \ner v_n \, \rho_n - \ner v_s ( \rho - \rho_s ) \nonumber \\
&=& \ner v_n \, \rho_n + \ner v_s \, \rho_s - \ner v_s \rho \; . 
\label{2.7}
\eea
Therefore, by comparing expression   (\ref{2.2})  with equations  (\ref{2.6}) and (2.7), one gets 
\be
 \left [ \ner P / V \right ]_{s} = \ner v_s \, \rho \; .
 \label{2.8}
\ee

\subsection{Energy density}

Let us now derive the value of the energy density $U/V$ for a portion of liquid in which both velocities $\ner v_n$ and $\ner v_s$ are nonvanishing. From equation (\ref{2.3}) it follows that the energy density  $\left [ U /V \right ]_{q.p.}$ of the quasi-particles gas is given by 
\be
\left [ U /V \right ]_{q.p.} = \int d \tau \, (\varepsilon - \ner v \ner p) \, n (\varepsilon - \ner v \ner p) \; , 
\label{2.9}
\ee
where $\ner v = \ner v_n - \ner v_s$.  A second order expansion in powers of $\ner v $ gives 
\be
\left [ U /V \right ]_{q.p.} \simeq \int d \tau \left [ \varepsilon \, n (\varepsilon ) + (\ner v \ner p)^2 {\der n (\epsilon )\over \der \varepsilon} + \mezzo (\ner v \ner p)^2  \varepsilon {\der^2 n \over \der \varepsilon ^2} \right ] \; . 
\label{2.10}
 \ee
By means of the replacement  $\ner p_i \ner p_j \rightarrow \delta_{ij} \, p^2/3$, one obtains  
 \be
 \left [ U /V \right ]_{q.p.} \simeq  U_0 /V  + v^2  \int d \tau \left \{  ( \varepsilon \, p^2 / 6)   {\der^2 n (\varepsilon ) \over \der \varepsilon ^2} - (p^2/3)  \left [ - {\der n (\varepsilon )\over \der \varepsilon} \right ]   \right \} \; ,  
 \label{2.11}
\ee
where 
\be
U_0 /V = \int d \tau \, \varepsilon \, n (\varepsilon )\; . 
\label{2.12}
\ee 
 Finally, in agreement with the expression (\ref{2.8}), the energy density $ \left [ U /V \right ]_s$ which is related to the condensate zero entropy motion  takes the form (in an   inertial reference system) 
 \be
  \left [ U /V \right ]_s = \mezzo \, \rho \, v_s^2 \;  . 
  \label{2.13}
  \ee 
Thus the total energy density is given by 
\be
U / V = U_0 / V +   \mezzo \, \rho \, v_s^2 +  \left (  \rho^*_n - \rho_n  \right )(\ner v_n - \ner v_s)^2 \; , 
\label{2.14}
\ee 
in which $\rho_n$ is defined in equation (\ref{2.5}) and 
\be
 \rho^*_n = \int d \tau \,   ( \varepsilon \, p^2 / 6)   {\der^2 n (\varepsilon ) \over \der \varepsilon^2} 
\; . 
\label{2.15}
\ee
   Expression (\ref{2.14}) can also be obtained by transforming the energy density of the liquid helium  from the inertial reference system in which $\ner v_s =0 $ to the inertial reference system in which $\ner v_s \not= 0$.
  
\subsection {Densities of free energy and entropy}
 
Equation (\ref{2.3}) implies that the   contribution $\left [ F /V \right ]_{q.p.}$  of the quasi-particles gas to the free energy density is given by 
\be
\left [ F /V \right ]_{q.p.} = kT \int d \tau \, \ln \left ( 1 - e^{-(\varepsilon - \ner v \ner p) / kT} \right ) \; , 
\label{2.16}
\ee
and the expansion up to  second order  in powers of the fluid velocities,  
$$
kT\int d \tau \, \ln \left ( 1 - e^{-(\varepsilon - \ner v \ner p) / kT} \right ) \simeq \int d \tau \, \Bigl \{kT  \ln \left ( 1 - e^{-\varepsilon / kT} \right ) 
 + \mezzo (\ner v \ner p)^2 \, {\der n (\varepsilon ) \over \der \varepsilon} \Bigr \} \; , 
$$
gives 
\be
\left [ F /V \right ]_{q.p.} = F_0/V - \mezzo \rho_n (\ner v_n - \ner v_s)^2 \; , 
\label{2.17}
\ee
where 
\be
F_0/V = kT \int d \tau \, \ln \left ( 1 - e^{-\varepsilon  / kT} \right ) \; . 
\label{2.18}
\ee
The free energy density $\left [ F /V \right ]_c$ which is associated with  the zero entropy motion of the condensate coincides with $ \left [ U /V \right ]_s = \mezzo \rho v_s^2$. 
Therefore, the density $F/V$ of free energy   of the liquid is given by 
\be
 F /V  = F_0/V + \mezzo \, \rho \, v_s^2  - \mezzo \rho_n (\ner v_n - \ner v_s)^2 \; . 
 \label{2.19}
 \ee
 By means of the thermodynamic relation  $F = U - TS$, the density of entropy turns out to be 
 \be
 T S / V =   \left ( U_0 - F_0 \right ) / V + \mezzo \left ( 2\rho^*_n - \rho_n \right ) (\ner v_n - \ner v_s)^2 \; .
 \label{2.20}
 \ee 

 \subsection{Mass densities}
 
This section contains the  theoretical determination of  the mass density $\rho^*_n$.  For completeness,  the computation \cite{L1, LL} of $\rho_n$ is also reported. In a neighbourhood of $p=0$, the Bose-Einstein distribution for quasi-particles 
  \be 
 n (\varepsilon )  = {1 \over e^{\varepsilon (p) / kT} - 1 } \;  ,   
\label{2.21}
\ee
can be approximated by 
 \be 
 n (\varepsilon )_{ph}   \simeq  {1 \over e^{u p / kT} - 1 } \;  ,   
\label{2.22}
\ee
and describes the phonons distribution. Whereas, in a neighbourhood of $p=p_0$, the distribution for the quasi-particles (rotons) can be approximated by the Maxwell-Boltzmann distribution 
\be 
 n (\varepsilon )_{r}   \simeq   e^{- \Delta / kT} \, e^{- (p - p_0)^2 / 2 m^* kT}  \;  ,     
\label{2.23}
\ee
because rotons constitute a low density gas. One can write 
\be
\rho_n =  \rho_{n, ph} + \rho_{n, r}\; , 
\label{2.24}
\ee
where  $\rho_{n,ph}$  reads 
\be
\rho_{n,ph} \simeq  \int {d^3p\over  h^3} {p^2 \over 3 kT} { e^{up/kT}\over \left ( e^{up/kT}-1\right )^2}= 
{2 \pi^2 (kT)^4 \over 45 \hbar^3 u^5} \; , 
  \label{2.25}
\ee
and  $\rho_{n,r}$  is given by 
\be
\rho_{n,r} \simeq  \int {d^3p\over  h^3} {p^2 \over 3 kT} \, e^{- \Delta / kT} \, e^{- (p - p_0)^2 / 2 m^* kT} \simeq { 2 p_0^4 \over 3  (2 \pi )^{3/2}  \hbar^3}\, \left ( m^*\over kT \right )^{1/2}   e^{-\Delta / kT}\; . 
\label{2.26}
\ee
Let us now concentrate on $ \rho^*_n$ shown in equation (\ref{2.15}); one can put 
\be
 \rho^*_n = \rho^*_{n, ph} +  \rho^*_{n, r} \; , 
\label{2.27}
\ee
in which the phonons contribution is given by 
\be
 \rho^*_{n,ph} \simeq  \int d \tau \,   ( p \, p^2 / 6u)   {\der^2 n (\varepsilon )_{ph} \over \der p^2} = {5\over 2} \, \rho_{n, ph} \; ,  
\label{2.28}
\ee
and the rotons part $ \rho^*_{n,r}$ turns out to be 
\bea
 \rho^*_{n,r} &\simeq& {e^{-\Delta / kT} \over 6 \, (kT)^2} \int d \tau \, p^2 \left [ \Delta + {(p-p_0)^2\over 2 m^*} \right ] \, e^{(p - p_0)^2 / 2 m^* kT} \nonumber \\
 &\simeq& \rho_{n, r }  \left [ {\Delta \over 2 kT} + {1 \over 4}    \right ] \; . 
\label{2.29} 
\eea 
 
\section{Rotating container}

Let us consider the case in which the container of the helium fluid is a  cylinder which is rotating around its axis with a constant angular velocity $\Omega$. The laboratory system is assumed to be an inertial reference system. The normal component of the liquid helium, which  has nonvanishing viscosity and interacts with the container walls, has perception of the motion of the container. Whereas  the superfluid component of liquid helium, with vanishing viscosity, is insensitive to the rotation of the vessel.  As a result,  after a transient period,  the whole system reaches the   stable condition in which, for small values of $\Omega$,  the viscous component of the fluid  is rotating with the same angular velocity of the container ($\ner v_n \not= 0$), whereas the superfluid component remains at rest ($\ner v_s =0$). 

In order to determine the precise motion of the quasi-particles gas which is induced by the rotation of the container, one can use the Landau reasoning  \cite{L1}. In the coordinate system which is rotating with the same angular velocity $\Omega$, the container is at rest, and the boundary conditions for the normal component of the liquid coincide with the stationary conditions of a static container.   
Therefore, in this reference system,  the motion is determined by the standard action principle and the statistical distribution is  expressed in terms of the  Gibbs factor  $\exp ( - E^\prime / kT)$, where $E^\prime$ denotes the energy of a quasi-particle in the rotating system  
\be
E^\prime = \varepsilon  - \ner \Omega \cdot (\ner r \wedge \ner p) \; .  
\label{3.1}
\ee
 Thus, in order to find the macroscopic motion of the quasi-particle gas, one can minimise the thermodynamic potentials which are obtained by means of the  energy  (\ref{3.1}), and this implies \cite{L1} that the quasi-particle gas is rotating as a whole with angular velocity $\Omega$.  

The same conclusion can also be obtained by considering the laboratory point of view, where   the equilibrium boundary condition  is determined by the requirement that  the part of the viscous liquid in contact with the walls of the container must have the same velocity of the walls. This implies that, in the equilibrium state, the viscous fluid must rotate as a solid body with the same angular velocity of the cylinder, so that there is no energy dissipation caused by friction.
 
To sum up, because of the nontrivial interactions  between the normal viscous component of the fluid with the moving walls of the container, in the laboratory system  the velocity $\ner v_n$ takes the value 
\be
\ner v_n =\ner v_n (\ner r ) = \ner \Omega \wedge \ner r  \; ,  
\label{3.2}
\ee
and, in agreement with Landau argument, the thermodynamic potentials can be computed 
by means of the standard rules of statistical mechanics  in which  the energy spectrum of the quasi-particles is given in equation (\ref{2.3}), with $\ner v_n$ shown in equation (\ref{3.2}) and $\ner v_s =0$.  

As the value of $\Omega $ increases, a critical value $\Omega_0$ is reached in which the condensate also starts moving ($\ner v_s \not= 0$). In order to proceed ---as much as possible--- according to an irrotational motion,   which means $\ner \nabla \wedge \ner v_s = 0$, the best solution consists in concentrating the vorticity in a single line (with a quantized vorticity).  This line must be closed, or it must have its end-points on the boundaries of the superfluid region. The stable configuration is obtained when a vortex line is created along the axis of the container. In the presence of a vortex line, the velocity $\ner v_s$ of the condensate   is directed as the tangent to  concentric circles belonging to a plane which is orthogonal to the axis of the cylinder and, for the minimum nontrivial value of the vorticity $2 \pi \chi = h / m$,  it has magnitude 
\be
 |\ner v_s | = v_s (r_\bot) = {\hbar \over m \, r_\bot } \; , 
\label{3.3}
\ee  
where $r_\bot$ denotes the distance from the central axis.  Now the main issue to  be discussed is the deduction of  the critical  value  $\Omega_0$.  

As a first possibility, one could try to extend the Landau reasoning, which is valid for the motion of the quasi-particles gas,  to  the condensate motion also. According to this hypothesis,  one should consider the energy $U_{vor}^\prime = U_{vor} - \ner \Omega \ner M_{vor}$, where $U_{vor}$ and $\ner M_{vor}$ denote the energy and the angular momentum ---in the laboratory system--- of the motion of the liquid in the presence of a vortex line.  The   minimisation of $U^\prime_{vor}$   leads \cite{LL,A,V} to the results: 
\begin{itemize}
\item the critical value of the angular velocity is given by  
\be
\overline \Omega_0  = {\hbar \over m R^2} \, \ln \left ( {R \over a} \right ) \; , 
\label{3.4}
\ee
where $R$ represents the radius of the cylinder and  $a$ denotes the size of the core of the vortex; 
\item the condensate starts moving in the same direction of the viscous normal component of the fluid, {\it i.e.} the velocity $\ner v_s$ is directed as $\ner v_n$ defined in equation (\ref{3.2}). 
\end{itemize}

 This procedure appears to be not completely established because the condensate displays no viscosity and is uninfluenced by the rotation  of the walls of the container. As a consequence, differently from the case of the normal viscous component of the fluid,  the boundary conditions for the condensate remain   the same in any rotating coordinate system independently of the specific value of the angular velocity.  So, as far as the motion of the condensate is concerned, it seems that the   thermodynamic potential to be minimised cannot be of the form $(U_{vor} - \ner \Omega \ner M_{vor})$,   because $\ner \Omega$ appearing in this expression   is totally undetermined since it is not fixed by the condensate boundary conditions. Also, expression (\ref{2.19}) shows that, if $\ner v_n$ and $\ner v_s$ have the same direction then, as a consequence of the formation of a vortex line,  the free energy of the system would increase;  this seems rather odd. 

\subsection {Critical angular velocity}

Let us consider then a second possibility, in which it is supposed that  the formation of the vortex line is induced by the motion of the quasi-particles gas. 

It is assumed that some external equipment  is acting on the system in order to maintain the temperature and the value $\Omega$ of the angular velocity fixed.  The velocity $\ner v_n$ of the normal component of the fluid is  specified in equation (\ref{3.2}). In this way,  the equilibrium boundary conditions between  the viscous component of the fluid and the walls of the rotating cylinder are satisfied, and the quasi-particle gas is indeed  in  thermal equilibrium.  The velocity $\ner v_s$  is the only variable we are interested  in;  this  variable specifies the (zero entropy) motion of the  frictionless superfluid component of helium.   In agreement with the laws of thermodynamics, it is assumed that the vortex line formation is determined by the minimisation condition of the free energy of the system. 

The free energy $F$ of the helium liquid  is obtained by integrating the density (\ref{2.19}) in the volume, 
\be
F = \int d^3 r \left \{ F_0 / V +   \mezzo \, \rho \, v_s^2 - \mezzo \rho_n  (\ner v_n - \ner v_s)^2 \right \} \; . 
\label{3.5}
\ee
The result is the  sum of three terms, 
\be
F = \widetilde F + F_I +  F_{II} \; . 
\label{3.6}
\ee
The first term $\widetilde F$ does not depend on $\ner v_s$, 
\be 
\widetilde F = F_0 -  \mezzo \int d^3r \,  \rho_n  |\ner v_n|^2  \; , 
\label{3.7} 
\ee
and then it is not involved in the computation of $\Omega_0$. The contribution $F_I$ is linear in $\ner v_s$, 
\be
F_I =  \int d^3r \,  \rho_n  \, \ner v_n \ner v_s  \; , 
\label{3.8} 
\ee
whereas $F_{II}$ is quadratic in $\ner v_s$, 
\be
F_{II} =  \mezzo \int d^3r \,   \rho_s \, |\ner v_s|^2 \; . 
\label{3.9}
\ee 
 The formation of the  vortex line takes place  
when $F_I + F_{II} < 0 $. For small velocities one can assume that the mass densities are constant; one finds 
 \be 
 F_I = \pm \, \rho_n { \pi  L R^2 \hbar  \over m} \, \Omega \; , 
 \label{3.10}
 \ee
\be
F_{II} =  \rho_s  { \pi L \hbar^2 \over m^2 } \, \ln \left ( {R \over a} \right ) \; , 
\label{3.11} 
\ee
where $ \pi R^2 L $ is the volume of the cylinder. The formation of the meniscus has been neglected because, for small velocities, 
it gives rise to minor effects.  The sign in expression (\ref{3.10}) is positive if the directions of $\ner v_n$ and $\ner v_s$ coincide, and it is negative when  $\ner v_n$ and $\ner v_s$ have opposite directions.   
Therefore the condition $F_I + F_{II} < 0 $ is satisfied when  

\begin{itemize}
\item $\Omega > \Omega_0$, in which  the critical value $\Omega_0$ of the angular velocity is given by  
\be
\Omega_0 =  \left ( {\rho_s \over \rho_n } \right ) \,  {\hbar \over m R^2} \, \ln   \left ( {R \over a} \right ) \; ; 
\label{3.12}
\ee

\item the condensate starts moving in the opposite direction of the viscous normal component of the fluid  ({\it i.e.} $\ner v_n \ner v_s = - |\ner v_n| \, | \ner v_s| < 0$).  

\end{itemize}

\noindent Expression (\ref{3.12}) looks similar to equation (\ref{3.4}) but predicts a nontrivial dependence of the critical angular velocity on the  temperature. In particular, $\Omega_0$ vanishes in the $T \rightarrow T_0$ limit, and tends to diverge when $T \rightarrow 0$. Perhaps, the result that $\ner v_n$ and $\ner v_s$ must have opposite directions  may appear  unexpected; in any case, this conclusion is also confirmed  by the requirement of thermodynamic stability, as it is shown in the next section.

\section{Thermodynamic relations}

The idea that  the motion of the condensate is  caused by the presence of the rotating gas of quasi-particles, and that the emergence of the vortex line is related to the minimisation of the free energy,   seems  to be quite reasonable. But of course only the comparison of the prediction  (\ref{3.12})   with the experiments will determine the actual reliability of this approach. 

In addition to the measure of the critical angular velocity  $\Omega_0$, one could also examine the behaviour  of certain  thermodynamic variables which are involved in the formation of the vortex line.   For fixed volume,  the equilibrium thermal states of helium II inside a rotating container can be characterised by the variables $\Omega $ and $T$. It is assumed that the quasi-particles gas is rotating with velocity $\ner v_n$ shown in equation (\ref{3.2}).  Let us consider  the critical curve $\Omega_0 (T)$ in the cartesian plane $(\Omega , T)$ shown in Figure~1.    The  curve $ \Omega_0 (T) $ describes  states of coexistence of two different types of liquid motions;  the points  in the region $(1)$ of the plane  correspond to states without vortex lines,   and the points in the region $(2)$ refer to states in which one vortex line is present. The transition from region $(1)$ to region $(2)$  corresponds to the formation of one vortex line.

 \vskip 0.8 truecm

\centerline {
\begin{tikzpicture} [scale=0.9] [>= triangle 60 ]
%
\draw [-> , line width=1pt ] (0,0) -- (0,5); 
\draw [-> , line width=1pt  ] (0,0) -- (10,0); 
\draw[line width=1pt, densely dashed ] (8,0) -- (8,4.5);
\draw [ line width=1pt ]  (8,0) .. controls (2,1.7) and (1.4,1.2) .. (1,4.5); 
\node at (-0.4,4.5) {$\Omega$};
\node at (9.7,-0.4) {$T$};
\node at (8.08,-0.4) {$T_0$};
\node at (2,0.7) {no vortex line};
\node at (1.2,1.3) {$(1)$};
\node at (3.5,2.7) {one vortex line};
\node at (3,3.3) {$(2)$};
\end{tikzpicture}
}

\vskip 0.5 truecm
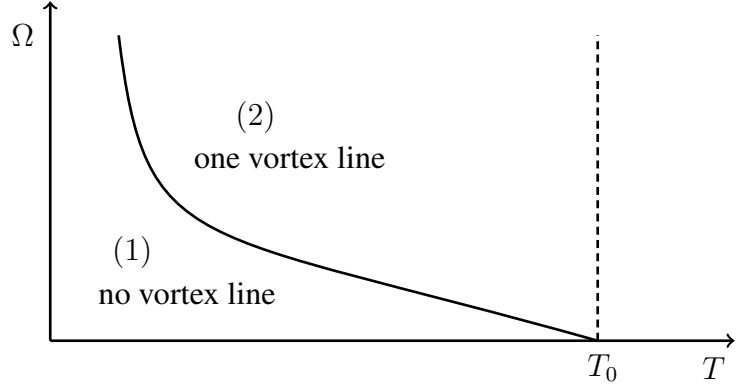
\centerline {{Figure~1.} Critical curve $\Omega_0 (T)$ in the $(\Omega , T)$-plane.}

\vskip 0.7 truecm

 \noindent   In crossing the critical curve,  the latent heat of vortex formation and the discontinuos change in the angular momentum  of the quasi-particles gas  determine the slope of the curve $\Omega_0 (T)$  by means of some kind of  Clapeyron equation. 

 In the differential of the free energy, $dF = - S dT - J d \Omega $, the variable  $J$ corresponds to the vertical component of the angular momentum of the quasi-particles gas. 
 Along the critical curve, the free energies $F_{(1)}$ and $F_{(2)}$ of the two types of  states are equal; therefore from equation $d F_{(1)} = dF_{(2)}$,  
 \be
 - S_{(1)} dT - J_{(1)} d \Omega_0 = - S_{(2)} dT - J_{(2)} d \Omega_0 \; ,  
 \label{4.1}
 \ee
one obtains 
\be
{ d \Omega_0 \over d T} = - {S_{(2)}  - S_{(1)} \over J_{(2)}- J_{(1)}}= - {\lambda \over T (J_{(2)} - J_{(1)} ) } \; ,  
\label{4.2}
\ee
where $\lambda = T (S_{(2)} - S_{(1)}) $ denotes the latent heat for the vortex formation.

\subsection{Angular momentum gap}

The total angular momentum of the liquid helium  is the sum of the angular momentum $\ner J$ of the quasi-particles gas, that for generic values of the velocities $\ner v = \ner v_n - \ner v_s$ is given by 
\bea
\ner J &=& \int d^3r \, d \tau \, n(\varepsilon - \ner v \ner p) \, \ner r \wedge \ner p   \; 
\simeq  \int d^3r \, d \tau \,  \ner r \wedge \ner p  \, (\ner v \ner p) \left [ - {\der n (\varepsilon ) \over \der \varepsilon }\right ] \nonumber \\ 
&\simeq &  \int d^3r \, \rho_n \, \ner r \wedge ( \ner v_n - \ner v_s ) \; , 
\label{4.3}
\eea
and the angular momentum $\ner M$ due to the motion of the condensate, 
\be 
\ner M = \int d^3r \, \rho \, \ner r \wedge \ner v_s \; . 
\label{4.4}
\ee 
The resulting total angular momentum is    
\be
\ner J + \ner M = \int d^3 r  \left ( \rho_s \, \ner r \wedge \ner v_s  + \rho_n \, \ner r \wedge \ner v_n \right ) \; . 
\label{4.5}
\ee
 When $\ner v_n = \ner \Omega \wedge \ner r$, with the angular velocity directed as the vertical axis $\ner \Omega = \Omega \widehat {\ner z }$,   from the expression (\ref{3.5}) of the free energy one gets 
\be
 {\der F \over \der \Omega } = -  \widehat {\ner z} \left (  \int d^3r \, \rho_n \, \ner r \wedge ( \ner v_n - \ner v_s )\right ) =  - \ner J_z \equiv - J \; . 
 \label{4.6}
  \ee
The discountinuous change of $J$, which is due to the formation of a vortex line, is given by 
\be
\Delta J = J_{(2)} - J_{(1)} =   - \widehat {\ner z} \left ( \int d^3r \, \rho_n \, \ner r \wedge \ner v_s \right ) =  \rho_n { \pi L R^2 \hbar \over m} \; . 
\label{4.7}
\ee 
It should be noted that, as a result of the formation of one vortex line, the vertical component of the angular momentum of the quasi-particle gas  increases, whereas the total angular momentum of helium II decreases   
\be
\Delta \left ( \ner J_z + \ner M_z \right ) = - \rho_s  { \pi L R^2 \hbar \over m} \; .
\label{4.8}
\ee

\subsection{Latent heat for vortex line formation}

The change in entropy due to the formation of a vortex line can be obtained by integrating  the change of entropy density (\ref{2.20}) in the volume,
\bea
\lambda = T (S_{(2)} - S_{(1)}) &=&  \mezzo \int d^3 r \, (2 \rho^*_n - \rho_n) (v_s^2 - 2 \ner v_n \ner v_s) \nonumber \\ 
&=&  \left (  2\rho^*_n  - \rho_n \right ) \left ( {\rho \over \rho_n } \right ) {\pi L \hbar^2 \over m^2}   \, \ln   \left ( {R \over a} \right ) \; . 
\label{4.9}
\eea
Equations (\ref{2.28}) and (\ref{2.29}) imply that the mass density $( 2\rho^*_n  - \rho_n ) $ which appears    in equation (\ref{4.9}) is positive, therefore after the formation of a vortex line   the value of the entropy is increased. The mass density $( 2\rho^*_n  - \rho_n ) $ is also related with the rate of increment  of  $\rho_n$ with the temperature. Indeed,  in the approximation in which the total mass density $\rho $ is constant, 
 from equation (\ref{3.12}) it follows 
 \be
 { d \Omega_0 \over d T} \simeq - {\rho \over \rho_n^2} \, {\der \rho_n \over \der T}
 \,  {\hbar \over m R^2} \, \ln   \left ( {R \over a} \right )\; . 
 \label{4.10}
 \ee
 By comparing equation (\ref{4.10}) with equations (\ref{4.2}), (\ref{4.7}) and (\ref{4.9}), one derives 
 \be
 T \, {\der \rho_n \over \der T} = 2 \rho^*_n - \rho_n \; . 
 \label{4.11}
 \ee
 Equation (\ref{4.11}) can also be obtained form expressions (\ref{2.19}) and (\ref{2.20}) 
 by means of the relation $S = - (\der F / \der T)_{V}$, or it can be derived directly from the definitions (\ref{2.5}) and (\ref{2.15}).   
 
The macroscopic motions of the liquid which are associated with the two  velocities $\ner v_n$ and $\ner v_s$ represent ``ordered''  motions of the elementary constituents of the fluid,   as opposed to the chaotic thermal motion of the atoms. In the case of a rotating container, a possible measure of the ordered motion of the fluid  is given by the magnitude of its total angular momentum. By keeping the value of $\Omega$ fixed, during the formation process   of a vortex line the amount of macroscopic ordered motion reduces, and the amount of disordered microscopic motion (value of entropy) grows. Precisely  because $\ner v_n$ and $\ner v_s$ have opposite orientations,   the discontinuous change of the entropy   is positive and the change in the total angular momentum is negative. 

\subsection {Numerical estimations}

Since $\Omega_0$ is obtained by minimising a quadratic function of the macroscopic  velocities of the liquid ---in which the value of $\ner v_s $ is specified in equation (\ref{3.3}) and $\ner v_n$ is described in equation (\ref{3.2})---, $\Omega_0 $ is proportional to $\overline \Omega_0$ shown in equation (\ref{3.4}).  The proportionality coefficient $D$, given by 
\be
D =  {\rho_s \over \rho_n} = {\rho \over \rho_n } - 1   \; , 
\label{4.12}
\ee
nontrivially depends on the temperature $T$. 
In the range $0.6 \leq T  \leq 2.1$ where the 
temperature is expressed in Kelvin, the rotons contribution to the mass density turns out to be  dominant \cite{LL,CA,Z}.  By using the  experimental data \cite{CA,Z} of the normal fluid ratio   $\rho_n / \rho $,  the resulting values of $D$ are shown in Table~1. 

\smallskip

\begin{center}

Table 1: Values of $D$ and of the normal fluid ratio at different temperatures.
\end{center}

\begin{center}
\begin{tabular}{|c|c|c|c|c|c|}
\hline 
$D$ & ~  & $2.34 \times 10^4$ & $1.03 \times 10^3$ & $1.31 \times 10^{2}$ &$33$ \\
 \hline
 $\rho_n / \rho $ & ~ & $4.27 \times 10^{-5}$ & $9.66 \times 10^{-4}$ & $7.52 \times 10^{-3}$ & $2.92 \times 10^{-2}$ \\ 
 \hline
 $T$ (K) & ~ &  0.6 & 0.8 & 1 & 1.2 \\
\hline
\hline
$D$ & $12$ & $4.88$ & $2.12$ & $0.78$ & $0.35$  \\
\hline
$\rho_n / \rho $ & $7.54 \times 10^{-2}$  & $0.17$ & $0.32$ & $0.56$ & $0.74$  \\
 \hline 
 $T$ (K) & 1.4 & 1.6 & 1.8 & 2.0 & 2.1 \\
 \hline
\end{tabular}
\end{center}

\smallskip

\noindent In the interval from $2.1$ to $0.8$ Kelvin, equation (\ref{3.12}) predicts a variation of $\Omega_0$ of three orders of magnitude.  This effect becomes even more important at lower temperatures.  For $T < 0.6 $~K, as the temperature become smaller the value of $D$ is rapidly increasing with the approximate behaviour  $\propto T^{-4}$. In the $ T \rightarrow 0 $ limit,  the asymptotic value of $D$  is given by 
\be
D \, \longrightarrow \,  {\rho \over  \rho_n} \simeq   \rho \, {45 \, \hbar^3 u^5 \over 2 \pi^2 (kT)^4 }\; , 
 \label{4.13}
\ee
where $\rho$ denotes the helium mass density, $ \rho \simeq 0.145$~$\hbox{g/cm}^3$.    The asymptotic behaviour (\ref{4.13}) is a consequence of the fact that  the density of quasi-particles vanishes in the $T \rightarrow 0 $ limit.    On the other hand, in a neighbourhood of the transition temperature $ T =T_0$, the low density approximation for the quasi-particles gas cannot be adopted. When $T \approx T_0$, in order to determine the value of the free energy of the system, the interactions between quasi-particles should be taken into account.       

\subsection{Vortices array}

Finally, when $\Omega \gg \Omega_0$ several vortices  are formed. The experimental data can be described as   

\begin{quote}

{\it Provided that the angular velocity $\Omega $ is not too small,   the vortex lines in uniformly rotating helium are straight and parallel to the axis of rotation, and they form an array with uniform density ... } \cite{FRN}   
\end{quote}

\noindent  Computations on the formation of vortex patterns in rotating superfluid, which are based on the minimisation of  $(U_{vor} - \ner \Omega \ner M_{vor})$, can be found for instance in Ref.\cite{CZ}.  

Differently from the case of a single vortex, the 
presence (and the time evolution) of a vortex array in the fluid  is not described by  stationary velocity fields for the fluid components. At any fixed time, the  spatial positions of the the vortex filaments break the continuous rotational symmetry around the vertical axis of the cylinder. The cores of the vortices posses nontrivial velocities that, combined with the localized positions of the filaments,  give rise to a nontrivial space and time dependence of the velocity field  $\ner v_s $, and then of the quasi-particles energy $E_v (\ner p )$.  Consequently, in order to describe the details of the emergence of a vortex array, one needs to consider the full set of thermodynamic and hydrodynamic equations (containing the relevant friction and drag parameters)  for the dynamics of liquid helium II. 

In addition to $\ner v_s$ and $\ner v_n$, the positions and the velocities of the vortex filaments must be specified.  If the position of one filament is  parametrised as $\ner s (\xi , t )$, where $\xi $ represents the  arclength, then the time evolution of  $\ner s$ can be approximated \cite{FRN,BSAM, KKR} by 
\be
{d \ner s \over dt} \simeq \ner v_s  + \alpha \, \ner s^\prime \wedge (\ner v_n - \ner v_s) - \alpha^\prime \, \ner s^\prime \wedge [ \ner s^\prime \wedge (\ner v_n - \ner v_s)] \; , 
\label{4.14}
\ee
  where $\ner s^\prime = d \ner s / d \xi $ and $\alpha$ and $\alpha^\prime $ denote the mutual friction coefficients. In the case of $n$ vortices, the velocities $\{ \ner v_s , \ner v_n , d \ner s_j / dt \} $, for $j = 1,2,...n $, display nontrivial space and time dependence and intricate couplings.   The cores of the vortices can be understood as defects in the condensate;     they  interact with the quasi-particles gas and are dragged by the rotating viscous fluid component.  As a result, the   superfluid vortex filaments tend to align with the normal fluid vorticity.  This effect has been observed, for instance, in the case of evolving turbulent flows, where the driven motion of the quantum vortex filaments by the normal fluid velocity has been determined and computed  \cite{KKR}  by means of numerical simulations.
  
It should be noted that, in the case of a rotating container, the formation of a single quantum vortex corresponding to a counter-rotating superfluid flow ---which is proposed in the present article---  is not in contradiction with the  behaviour of the motion of the vortex filaments. The emergence of one  quantum straight vortex line, which is placed on the axis of the rotating cylinder, differs from the  dragged motion of the core of the vortex filaments of an array because, unlike the dynamics of the quantum filaments, the  nucleation of the first vortex line  is specified by the minimisation condition of the  free energy (\ref{3.6}), as discussed in Section~3. 
  
\section{Conclusions}

In this article it has been proposed that, with a rotating container, the formation of  the first vortex lines  of the superfluid component of helium II  is caused by  the presence of the rotating quasi-particles gas, and that the critical angular velocity  $\Omega_0$ for the formation of one vortex  line can be obtained by minimising the free energy of the system. During the emergence of the first vortex line, the condensate starts moving in the opposite direction with respect to the motion of the rotating viscous component of the liquid, the entropy of the system increases and the total angular momentum decreases.   The  value of $\Omega_0$   that has been derived displays a nontrivial dependence on the  temperature;  $\Omega_0(T)$ vanishes in the $T \rightarrow T_0$ limit, and tends to diverge when $T \rightarrow 0$. Numerical estimations of the behaviour of $\Omega_0 (T) $ as a function of the temperature have been presented. 
It has been shown that the latent heat for the formation of one vortex line  and the corresponding discontinuos change in the angular momentum  of the quasi-particles gas  determine the slope of the curve $\Omega_0 (T)$  through  a sort of Clapeyron equation. The increment in the entropy and the   reduction of the total angular momentum of the liquid during the vortex formation have been determined. 

As far as  the experimental side is concerned, the direct determination of the condensate circulation of the first nucleated vortex appears to be  difficult to implement.  Presumably, it will be more easy to observe the consequences of the counter-rotating flow, like for instance the nontrivial temperature dependence of $\Omega_0(T)$  or the presence of the latent heat for the vortex formation.   
The new  technological  developments \cite{SZZ} in the study of cold atoms condensates will probably permit to measure some of the effects of the anti-correlation between the velocities of the superfluid and normal components. In the case of cold atoms condensates, the complete control on the boundary conditions for the viscous component of the fluid is crucial to test the mechanism discussed in this article.

\end{document}